%% file: text.tex
\RequirePackage{lineno}

\documentclass[aps,prd,twocolumn,superscriptaddress,showpacs]{revtex4}

\usepackage{graphicx}
\usepackage{dcolumn}
\usepackage{bm}
\usepackage{color}
\usepackage{subfigure}

\newcommand{\BR}{\ensuremath{\mathcal B}}

\begin{document}
\title{Measurement of the branching fraction ${\mathcal{B}}(\Lambda^0_b\rightarrow \Lambda^+_c\pi^-\pi^+\pi^-)$  at CDF}
\input{September2011_Authors}
\date{\today} 
\begin{abstract}
We report an analysis of the $\Lambda^0_b\rightarrow\Lambda^+_c\pi^-\pi^+\pi^-$
decay  in a data
sample collected by the CDF II detector at the Fermilab Tevatron 
corresponding to 2.4 fb$^{-1}$ of integrated luminosity.  
We reconstruct the currently largest samples of
the decay modes
$\Lambda^0_b\rightarrow \Lambda_c(2595)^+\pi^-$ 
(with $\Lambda_c(2595)^+\rightarrow\Lambda^+_c\pi^+\pi^-$),
$\Lambda^0_b\rightarrow \Lambda_c(2625)^+\pi^-$
(with $\Lambda_c(2625)^+\rightarrow\Lambda^+_c\pi^+\pi^-$),
$\Lambda^0_b\rightarrow \Sigma_c(2455)^{++}\pi^-\pi^-$ 
(with $\Sigma_c(2455)^{++}\rightarrow\Lambda^+_c\pi^+$),
and $\Lambda^0_b\rightarrow \Sigma_c(2455)^0\pi^+\pi^-$
(with $\Sigma_c(2455)^0\rightarrow\Lambda^+_c \pi^-$)
and measure the branching fractions relative to the
$\Lambda^0_b\rightarrow\Lambda^+_c\pi^-$
branching fraction. We measure the ratio 
$\BR(\Lambda^0_b\rightarrow \Lambda^+_c\pi^-\pi^+\pi^-)
/\BR(\Lambda^0_b\rightarrow \Lambda^+_c \pi^-)
= 3.04 \pm 0.33(\text{stat})^{+0.70}_{-0.55}(\text{syst})$
which is used to derive
$\BR(\Lambda^0_b\rightarrow \Lambda^+_c\pi^-\pi^+\pi^-)= 
(26.8^{+11.9}_{-11.2})\times10^{-3}$.
 \end{abstract}
  
  \pacs{14.20.Mr 14.20.Lq}  
  
  
  \maketitle
  
  

\section{Introduction}
Due to the high $b$-quark mass, weak decays of baryons containing a $b$ quark are a good
testing ground of some approximations in quantum chromodynamics (QCD) 
calculations, such as heavy-quark effective theory (HQET) \cite{HQET}.
Alternatively, when using such calculations, 
the $\Lambda^0_b$ may provide a determination 
of the Cabibbo-Kobayashi-Maskawa (CKM) couplings
with systematic uncertainties different from the 
determinations from the decays of $B$ mesons \cite{CKM_Theory}.
While the $B$ mesons are well studied, less is known
about the $\Lambda^0_b$ baryon.
Only nine decay modes
of the $\Lambda^0_b$ have been observed so far, with the sum of their 
measured branching fractions of the order of only 0.1 and with large 
uncertainties on the measurements \cite{PDG2010}.
While theoretical predictions are available for the
$\Lambda^0_b\rightarrow\Lambda^+_c\pi^-$ branching
fraction \cite{Lc1pi_Theory}, 
no prediction is currently available for the  
$\Lambda^0_b\rightarrow\Lambda^+_c\pi^-\pi^+\pi^-$
decay mode.
LHCb recently reported the measurement of the ratio of
branching fractions
$\BR(\Lambda^0_b\rightarrow \Lambda^+_c \pi^-\pi^+\pi^-)
/\BR(\Lambda^0_b\rightarrow \Lambda^+_c \pi^-)$
$= 1.43 \pm 0.16(\text{stat})\pm 0.13(\text{syst})$
\cite{LHCb}.

This paper reports a study of the 
$\Lambda^0_b\rightarrow\Lambda_c^+\pi^-\pi^+\pi^-$ decay mode
and is especially distinguished by the high yields and
high precision measurement of the 
$\Lambda^0_b\rightarrow\Lambda_c^+\pi^-\pi^+\pi^-$
resonant contributions, the following decay modes:
\begin{description}
\item $\Lambda^0_b\rightarrow \Lambda_c(2595)^+\pi^-$,
\item $\Lambda^0_b\rightarrow \Lambda_c(2625)^+\pi^-$,
\item $\Lambda^0_b\rightarrow \Sigma_c(2455)^{++}\pi^-\pi^-$,
\item $\Lambda^0_b\rightarrow \Sigma_c(2455)^0\pi^+\pi^-$.
\end{description}
We measure the branching fraction of each resonant 
decay mode relative to the 
$\Lambda^0_b\rightarrow\Lambda^+_c\pi^-$ decay mode, 
and the ratio of branching fractions
$\BR(\Lambda^0_b\rightarrow \Lambda^+_c\pi^-\pi^+\pi^-)
/\BR(\Lambda^0_b\rightarrow \Lambda^+_c\pi^-)$. 
The  measurement is performed using a sample of $p\overline{p}$
collisions corresponding to 2.4 fb$^{-1}$ integrated luminosity 
collected by CDF~II
between February 2002 and May 2007. 
We reconstruct $\Lambda^0_b$ decays from particles whose trajectory projections 
in the plane transverse to the beamline do not intersect the beamline (displaced tracks). 
The signal yields of interest are extracted by fitting 
mass differences to minimize the effect of systematic uncertainties.
As a crosscheck, we repeat the analysis on the reference decay
modes $B^0\rightarrow D^-\pi^+\pi^-\pi^+$ and $B^0\rightarrow D^-\pi^+$.

The structure of the paper is as follows.
Section~\ref{sec:detector} describes the detector systems
relevant to this analysis.
Event selection and $\Lambda^0_b\rightarrow\Lambda^+_c\pi^-\pi^+\pi^-$
and $\Lambda^0_b\rightarrow\Lambda^+_c\pi^-$ candidate
reconstruction are described in Sec.~\ref{sec:eventreco}.
In Sec.~\ref{sec:yields} we present the signal yields.
In Sec.~\ref{sec:measurement} we describe the evaluation of the detector acceptance
and  the relative branching fraction measurements, while in 
Sec.~\ref{sec:systematics} the systematic uncertainties are discussed.
Final results are reported in Sec.~\ref{sec:results}.

\section{The CDF II Detector and Trigger}
\label{sec:detector}
  The CDF II detector is a multipurpose magnetic spectrometer surrounded by
  calorimeters and muon detectors. 
The components relevant
  to this analysis are briefly described here. A more detailed 
  description can be found elsewhere~\cite{CDF_Acosta}.
  A silicon microstrip detector (SVX and ISL)~\cite{Silicon_Sill} 
  and a cylindrical drift chamber
  (COT)~\cite{COT_Affolder} 
immersed in a 1.4 T solenoidal magnetic field
  allow the reconstruction of charged particle trajectories in 
the pseudorapidity \cite{Pseudorapidity} 
range  $|\eta| < 1.0$ \cite{CDF-coordinates}. 
  The SVX detector consists of microstrip sensors arranged in six cylindrical
shells around the beamline with radii between 1.5 and 10.6 cm, and with a total $z$ coverage of 90 cm. 
The first SVX layer, also referred to as the L00 detector, is made of single-sided 
sensors mounted on the beryllium beam pipe. The remaining five SVX  layers are made 
of double-sided sensors and  divided into three contiguous five-layer sections along 
the beam direction $z$. The two additional silicon layers of the ISL help to link tracks 
in the COT to hits in the SVX.
    The COT has 96 measurement layers between 40 and 137 cm in radius, 
  organized into 
  alternating axial and $\pm 2^{\circ}$ stereo superlayers.
 The charged particle transverse momentum resolution is 
$\sigma_{p_{T}}/p_{T} \simeq 
  0.07\%\, p_{T}$ (GeV/$c$), 
and the resolution on the transverse distance of closest approach of the particle trajectory to the beamline
 (impact parameter, $d_0$) is $\approx$40 $\mu$m,
 including a $\approx$30 $\mu$m contribution from the beamline.
  
 
  Candidate events for this analysis are selected by a three-level on-line event selection 
  system (trigger).  At level 1, charged particles are reconstructed in the
  COT axial superlayers by a hardware processor, the Extremely Fast
  Tracker (XFT)~\cite{XFT_Thomson}. Two
  charged particles are required with transverse momenta $p_T \geq 2$ GeV/$c$.
At level 2, the Silicon Vertex Trigger (SVT) \cite{SVT_Ashmanskas} associates
  SVX  $r-\phi$ position measurements with XFT tracks.  This provides a precise 
  measurement of the track impact parameter $d_0$.
We select $b$-hadron candidates 
  by requiring two SVT tracks with 120 $\mu$m
  $\leq d_0 \leq$ 1000 $\mu$m. To reduce background from light-quark
  jet pairs, the two trigger tracks are required to have an opening
  angle in the transverse plane $2^{\circ} \leq \Delta\phi \leq
  90^{\circ}$. The tracks must also satisfy the requirement $L_{T} > 200$ $\mu$m,
  where $L_{T}$ is defined as the distance in the transverse plane
  from the beam line to the two-track intersection point, projected onto the
  two-track momentum vector. The level 1 and 2 trigger requirements are then confirmed at trigger level 3, where the event is fully reconstructed.
    
\section{Event Reconstruction}
\label{sec:eventreco}
The search for $\Lambda^0_b\rightarrow\Lambda^+_c\pi^-\pi^+\pi^-$ 
and 
$\Lambda^0_b\rightarrow\Lambda^+_c \pi^-$ 
candidates begins with the 
reconstruction of the $\Lambda^+_c$ using the three-body decay 
$\Lambda^+_c\rightarrow p K^-\pi^+$ \cite{chrg_cnj}. 
Three tracks, assumed to 
be a kaon, a proton, and a pion, with a total charge of +1,
are fit to a common vertex.
No particle identification is used in this analysis. All particle
hypotheses consistent with the candidate decay chain 
are considered.
Additional selection criteria (cuts) are applied on fit
probability ($P(\chi^2(\Lambda^+_c))>10^{-5}$), 
transverse momentum ($p_T(\Lambda^+_c) > 4.0$ GeV/c), 
and transverse decay length relative to the beamline 
($L_T(\Lambda^+_c) > 200$ $\mu$m).
We also require $p_T(p)>p_T(\pi^+)$, to suppress
random-track combinatorial background. 
The reconstructed $\Lambda^+_c$ mass ($m(\Lambda^+_c)$)
distribution is
comparable to the one reported in Ref.~\cite{prd_cbaryons}.
The reconstructed $\Lambda^+_c$ mass is required to be close 
to the known $\Lambda^+_c$ mass 
(2.240 - 2.330 GeV/c$^2$) \cite{PDG2010}.
Since mass differences are used to search for 
the resonances, no mass constraint is applied
in the $\Lambda^+_c$ reconstruction.  
The $\Lambda^0_b\rightarrow\Lambda^+_c\pi^-\pi^+\pi^-$ 
($\Lambda^0_b\rightarrow\Lambda^+_c \pi^-$)
candidate is reconstructed by
performing a fit to a common vertex of the 
reconstructed $\Lambda^+_c$ and  three (one) 
additional tracks, assumed to be pions, with $p_T>0.4$ GeV/c, 
and a total charge of --1. 
For all the possible track pairs out of the six (four) tracks 
that form the $\Lambda^0_b$ candidate, we require the difference
between the $z$ coordinate of the points of closest approach
of the two tracks to the beam to be less than 5 cm.
Additional cuts on the $\Lambda^0_b$ 
candidate fit probability ($P(\chi^2(\Lambda^0_b))>10^{-4}$), 
transverse momentum ($p_T(\Lambda^0_b) > 6.0$ GeV/c), 
transverse decay length relative to the beamline 
($L_T(\Lambda^0_b) > 200$ $\mu$m), 
and $\Lambda^+_c$ transverse decay length relative to the beamline 
($L_T(\Lambda^+_c) > 200$ $\mu$m)
and to the $\Lambda^0_b$ vertex
($L_T(\Lambda^+_c$ from $\Lambda^0_b) > -200$ $\mu$m) are applied.
We also require that the transverse momentum of the pion
produced in the $\Lambda^+_c$ decay is larger than the
transverse momentum of the same-charge pion produced in
the $\Lambda^0_b$ decay, which considerably reduces the combinatorial
background due to the larger boost of the pion 
produced in the $\Lambda^+_c$ decay.
To improve the purity of the 
$\Lambda^0_b\rightarrow\Lambda^+_c\pi^-\pi^+\pi^-$ 
signal, we optimize the analysis cuts to
maximize the signal significance
${\mathcal{S}}/\sqrt{{\mathcal{S+B}}}$.
The number of 
$\Lambda^0_b\rightarrow\Lambda^+_c\pi^-\pi^+\pi^-$ candidates 
${\mathcal{S}}$ and the number of background events
${\mathcal{B}}$ are estimated
in data by performing a fit of the $m(\Lambda^0_b)$ distribution.
This procedure determines the final selection criteria:
$p_T(\Lambda^0_b)>9.0$ GeV/c, 
$L_T(\Lambda^0_b)/\sigma_{L_T(\Lambda^0_b)}>16$,
$d_0(\Lambda^0_b)< 70$ $\mu$m, and $\Delta R(\pi^-\pi^+\pi^-) < 1.2$,
where $d_0(\Lambda^0_b)$ is the impact parameter of the
reconstructed $\Lambda^0_b$ candidate relative to the beamline 
and $\Delta R(\pi^-\pi^+\pi^-)$ is the maximum
$\sqrt{\Delta\eta^2+\Delta\phi^2}$ distance between the two pions
in each of the three possible pairs of pions. 
We verified that by splitting the data sample in two independent
samples, the optimization procedure yields the same final selection
criteria when applied separately to the two samples, and that the 
$\Lambda^0_b\rightarrow\Lambda^+_c\pi^-\pi^+\pi^-$ yield is evenly 
distributed. This ensures that our optimization procedure
does not introduce a bias on the branching fraction measurement.
To reduce possible systematic effects in the
estimate of the reconstruction efficiency due
to Monte Carlo simulation model inaccuracy, the same selection cuts optimized 
for $\Lambda^0_b\rightarrow\Lambda^+_c\pi^-\pi^+\pi^-$ are also applied
to the selection of the $\Lambda^0_b\rightarrow\Lambda^+_c\pi^-$
signal, except for the $\Delta R(\pi^-\pi^+\pi^-)$ cut.

\section{Determination of the Signal Yields}
\label{sec:yields}
\begin{figure}[htbp] 
\begin{center}
\subfigure[]{\label{mass_a}
\includegraphics[width=8cm]{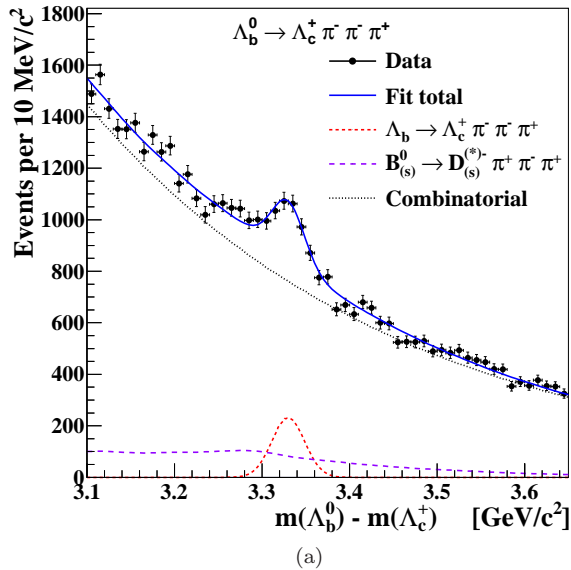}
}
\subfigure[]{\label{mass_b}
\includegraphics[width =8cm]{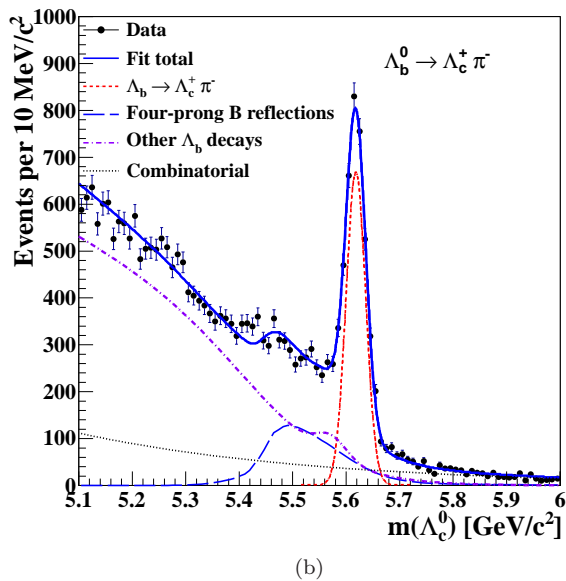}
}
\caption{The reconstructed invariant mass spectra 
after applying all the selection criteria: 
(a) the mass difference $m(\Lambda^0_b)-m(\Lambda^+_c)$ distribution 
of the $\Lambda^0_b\rightarrow\Lambda^+_c \pi^- \pi^+ \pi^-$ candidates;
(b) the $m(\Lambda^0_b)$ distribution of the 
$\Lambda^0_b\rightarrow\Lambda^+_c \pi^-$ candidates.}
\label{mass}
\end{center}
\end{figure}

Figure\ \ref{mass}\subref{mass_a} shows the
distribution of the difference between the reconstructed
$\Lambda^0_b$ and $\Lambda^+_c$ masses, 
$m(\Lambda^0_b)-m(\Lambda^+_c)$, of the selected 
$\Lambda^0_b\rightarrow\Lambda^+_c\pi^-\pi^+\pi^-$ candidates with the
fit projection overlaid. 
A significant signal of $\Lambda^0_b\rightarrow\Lambda^+_c\pi^-\pi^+\pi^-$
is visible centered approximately at 3.330 GeV/c$^2$. 
Backgrounds include misreconstructed
multibody $b$-hadron decays (physics background) and random combinations
of charged particles that accidentally meet the selection requirements
(combinatorial background). We use an unbinned extended
maximum-likelihood fit
to estimate the $\Lambda^0_b\rightarrow\Lambda^+_c\pi^-\pi^+\pi^-$
signal yield. The signal peak is modeled with a Gaussian,
with mean and width left floating in the fit. 
The combinatorial background is modeled with an 
exponential function of $m(\Lambda^0_b)-m(\Lambda^+_c)$
with floating slope and normalization. 
The distribution of the main physics backgrounds, due to 
the $B^0_{(s)}\rightarrow D_{(s)}^{(*)-}\pi^+\pi^-\pi^+$ decay 
modes, are derived from simulation and included in the fit
with fixed shape and floating normalization.
The $\Lambda^0_b\rightarrow\Lambda^+_c\pi^-\pi^+\pi^-$
yield estimated by the fit of the data is 1087$\pm$101
candidates, the world's largest sample currently available
of this decay mode. 
Figure\ \ref{mass}\subref{mass_b} shows
the $\Lambda^0_b$ mass distribution of the
selected $\Lambda^0_b\rightarrow\Lambda^+_c\pi^-$ 
candidates. 
The $\Lambda^0_b$ mass distribution
is described by several components: 
the $\Lambda^0_b\rightarrow\Lambda^+_c\pi^-$ Gaussian signal,
a combinatorial background, 
reconstructed $B$ mesons that pass the $\Lambda^+_c\pi^-$
selection criteria, partially reconstructed $\Lambda^0_b$
decays ({\em e.g.} $\Lambda^0_b\rightarrow\Lambda^+_c l^- \bar{\nu}_l$), 
and fully reconstructed $\Lambda^0_b$ decays
other than $\Lambda^+_c\pi^-$ 
({\em e.g.} $\Lambda^0_b\rightarrow\Lambda^+_c K^-$).
Also in this case the distributions of physics backgrounds 
are derived from simulation and included in the fit with fixed
shapes and floating normalization, as detailed in Ref.~\cite{prl_Lb_life}.
The $\Lambda^0_b\rightarrow\Lambda^+_c\pi^-$ yield estimated by the 
fit of the data is 3052$\pm$78 candidates.

\begin{figure*}[htbp] 
   \centering
\subfigure[]{\label{fig:resonances_1a}
\includegraphics[width=8cm]{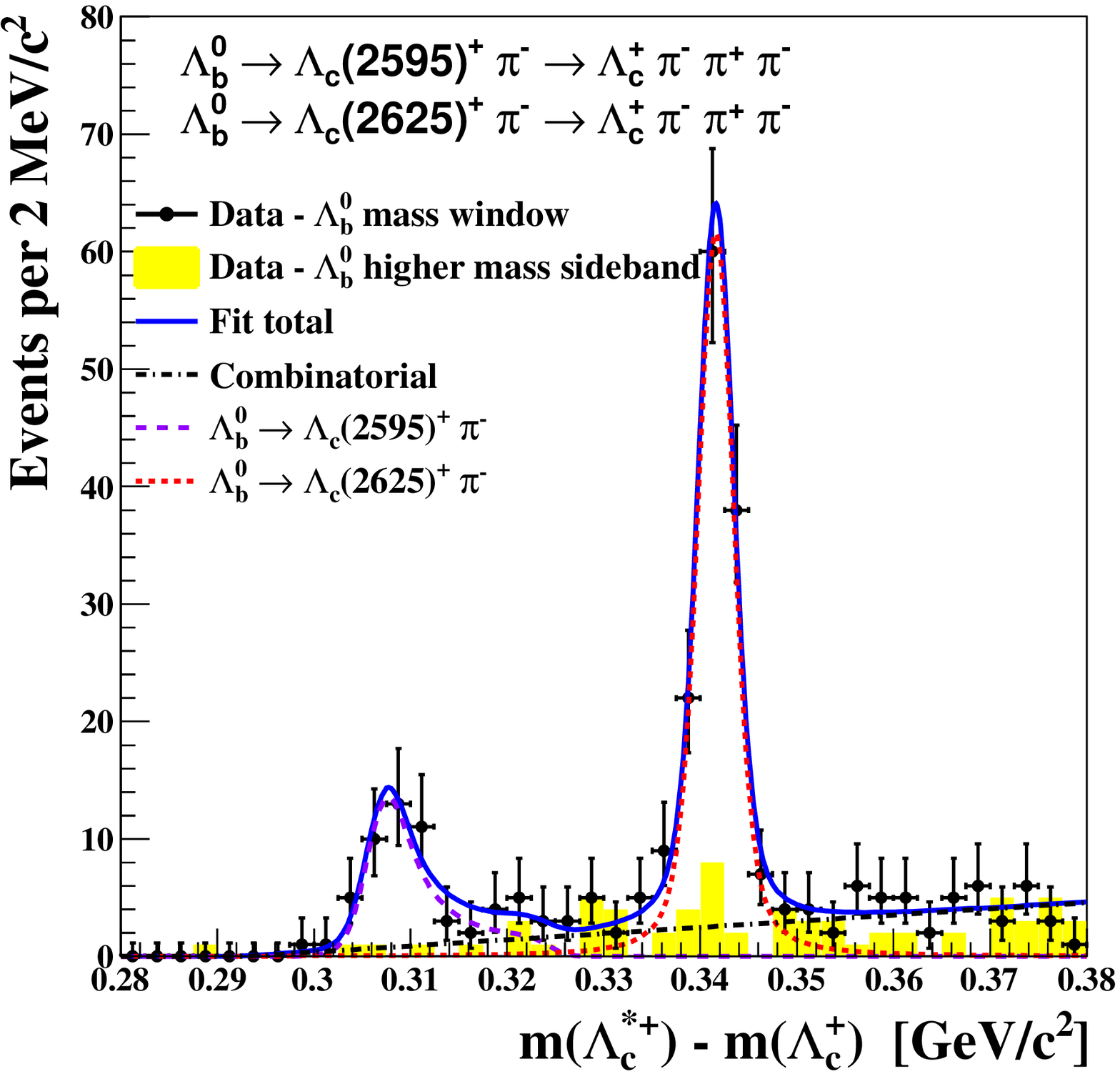}
}
\subfigure[]{\label{fig:resonances_1b}
\includegraphics[width=8cm]{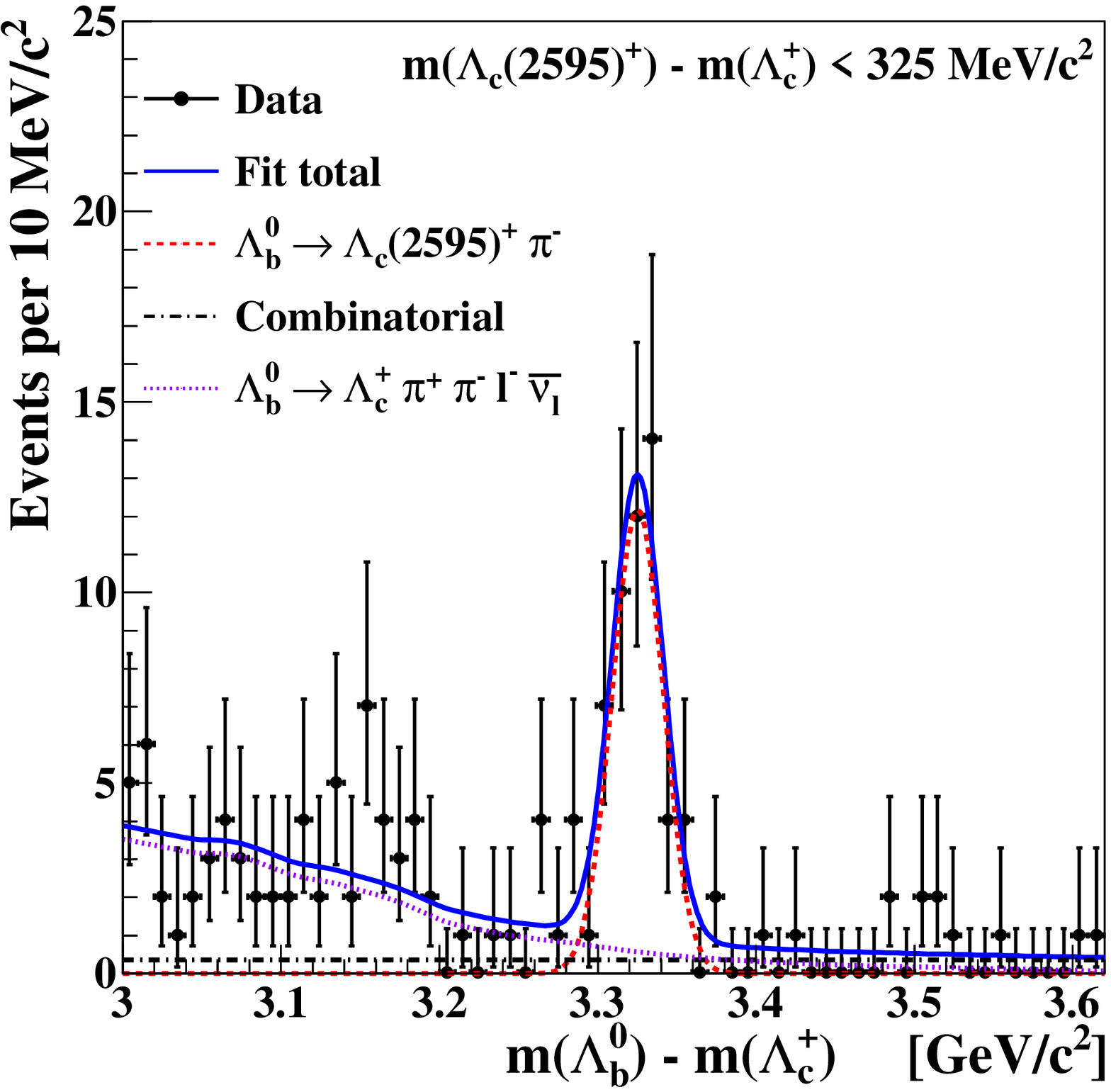}
}
\subfigure[]{\label{fig:resonances_1c}
\includegraphics[width=8cm]{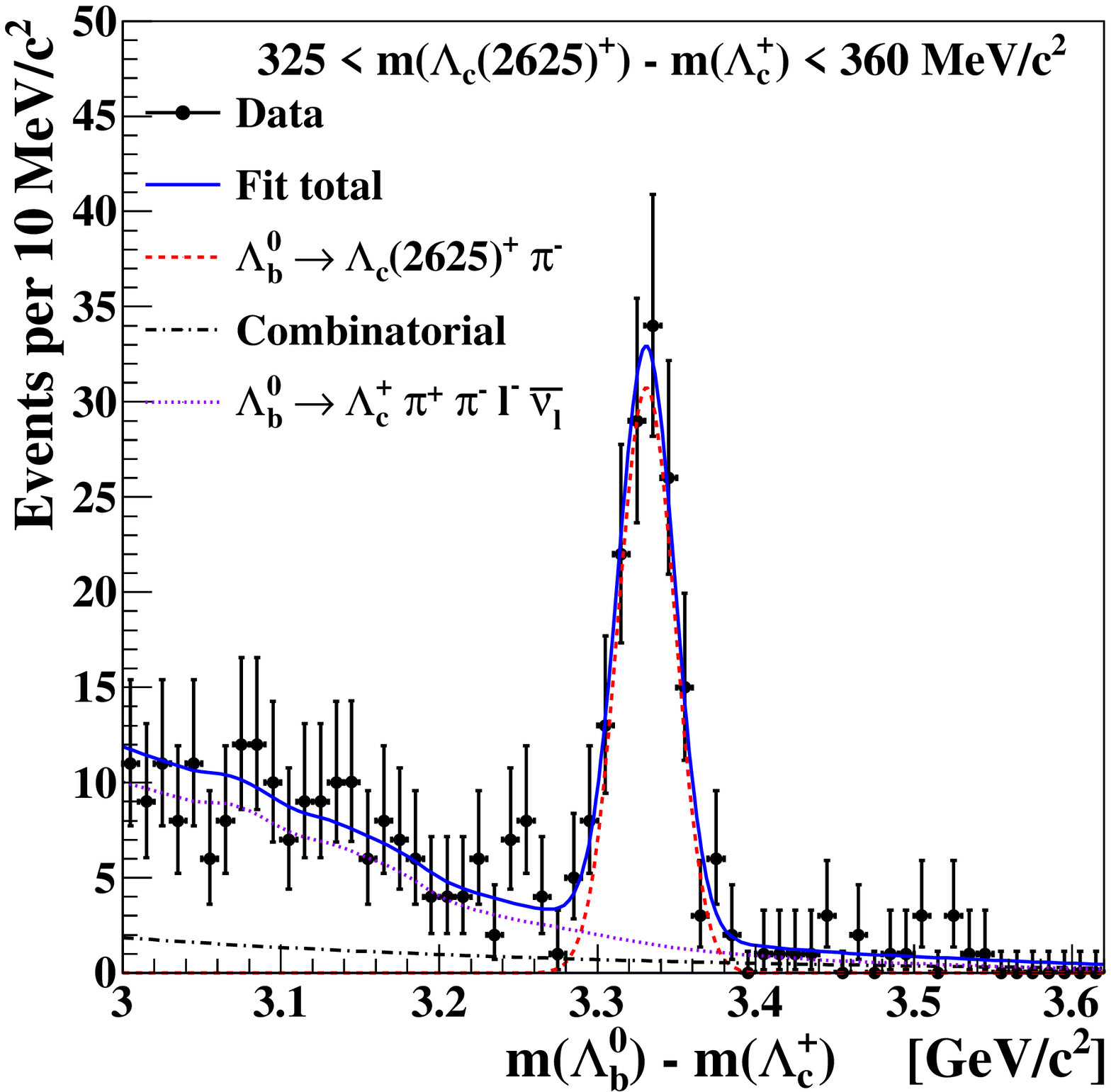}
}
 \caption{The $\Lambda^0_b\rightarrow\Lambda_c(2595)^+\pi^-$
and $\Lambda^0_b\rightarrow\Lambda_c(2625)^+\pi^-$ signals:
(a) $m(\Lambda_c^{*+})-m(\Lambda^+_c)$ distribution 
for candidates in a
$\pm3\sigma$ range ($\pm$57 MeV/c$^2$) around the $\Lambda^0_b$ mass;
(b) $m(\Lambda^0_b)-m(\Lambda^+_c)$ distribution restricted to candidates
in the region $m(\Lambda_c^{*+})-m(\Lambda^+_c)< 0.325$ GeV/c$^2$;
(c)  $m(\Lambda^0_b)-m(\Lambda^+_c)$ distribution restricted to candidates
in the region 0.325 
$< m(\Lambda_c^{*+})-m(\Lambda^+_c)< 0.360$ GeV/c$^2$.
}
\label{fig:resonances_1}
\end{figure*}

In the reconstructed $\Lambda^0_b\rightarrow\Lambda^+_c\pi^-\pi^+\pi^-$
sample we searched for the  resonant decay modes:
$\Lambda^0_b\rightarrow\Lambda_c(2595)^+\pi^-$,
$\Lambda^0_b\rightarrow\Lambda_c(2625)^+\pi^-$,
$\Lambda^0_b\rightarrow\Sigma_c(2455)^{++}\pi^-\pi^-$, 
and $\Lambda^0_b\rightarrow\Sigma_c(2455)^{0}\pi^+\pi^-$.
The available energy transferred to the decay products
in the decays of the charmed baryons ($\Lambda_c(2595)^+$,
$\Lambda_c(2625)^+$, $\Sigma_c(2455)^{++}$, and $\Sigma_c(2455)^0$)
into $\Lambda^+_c$ is small.
Therefore the differences of the reconstructed masses  
$m(\Lambda_c^{*+})-m(\Lambda^+_c)$,
$m(\Sigma_c(2455)^{++})-m(\Lambda^+_c)$, and
$m(\Sigma_c(2455)^0)-m(\Lambda^+_c)$ are determined with better
resolution than the masses of the charmed baryons, 
since the mass resolution of the $\Lambda^+_c$ signal
and most of the mass systematic uncertainties cancel 
in the difference.
Figure\ \ref{fig:resonances_1}\subref{fig:resonances_1a}
shows the $m(\Lambda_c^{*+})-m(\Lambda^+_c)$
distribution,
for $\Lambda^0_b\rightarrow\Lambda^+_c \pi^- \pi^+ \pi^-$ candidates
with mass in a $\pm 3\sigma$ range ($\pm 57$~MeV/c$^2$)
around the $\Lambda^0_b$ mass. The $\Lambda_c(2595)^+$
and $\Lambda_c(2625)^+$ signals are clearly visible.
Although there are two possible $\Lambda_c^{*+}$ 
candidates for each $\Lambda^0_b\rightarrow\Lambda^+_c\pi^-\pi^+\pi^-$
decay, only the candidate 
made with the $\pi^-$ with lower  $p_T$ 
has a value of $m(\Lambda_c^{*+})-m(\Lambda^+_c)$ in the
mass region where the
$\Lambda_c(2595)^+$ and $\Lambda_c(2625)^+$ signals are expected.
The $\Lambda_c(2595)^+$ and $\Lambda_c(2625)^+$
signal yields are estimated 
with an unbinned extended maximum-likelihood fit.
The $\Lambda_c(2595)^+$ and $\Lambda_c(2625)^+$
signals are modeled with two
non-relativistic Breit-Wigner functions 
convolved with the same Gaussian resolution function,
since the mass difference between the two resonances
is tiny. The background is modeled by a linear function. 
The $\Lambda_c(2595)^+$ natural width is mass
dependent to take into account the threshold
effects, as reported in Ref.~\cite{prd_cbaryons}, 
the $\Lambda_c(2625)^+$ natural width
and the width of the Gaussian resolution function are  free
parameters of the fit.
Table \ref{tab:yields} reports the estimated signal yields
and significances,
evaluated by means of the likelihood ratio test, $LR \equiv L/L_{bck}$,
where $L$ and $L_{bck}$ are the likelihood
of the signal  and no signal hypotheses, respectively \cite{LR}.

Figures~\ref{fig:resonances_1}\subref{fig:resonances_1b}
and \ref{fig:resonances_1}\subref{fig:resonances_1c}
show the $m(\Lambda^0_b)-m(\Lambda^+_c)$ distribution
restricted to candidates  with
$m(\Lambda_c^{*+})-m(\Lambda^+_c)< 0.325$ GeV/c$^2$
and 0.325$< m(\Lambda_c^{*+})-m(\Lambda^+_c)< 0.360$ GeV/c$^2$, respectively, 
i.e. compatible with the $\Lambda_c(2595)^+$ and $\Lambda_c(2625)^+$
expected signals. 
Each signal is modeled with a Gaussian function, with floating mean and width. 
The combinatorial background is modeled with an 
exponential function with floating slope and normalization, 
and the physics background, which is mainly due to semileptonic 
$\Lambda^0_b\rightarrow\Lambda^+_c\pi^-\pi^+ l^-\overline{\nu}_l$ decays,
is derived from simulation and introduced in the fit with fixed
shape and floating normalization.
We verified that the 
$\Lambda^0_b\rightarrow\Lambda_c(2595)^+\pi^-$ and 
$\Lambda^0_b\rightarrow\Lambda_c(2625)^+\pi^-$ yields
estimated by fitting the $m(\Lambda^0_b)-m(\Lambda^+_c)$ 
distributions are compatible with the yields reported 
in Table~\ref{tab:yields} with lower significance.

\begin{figure*}[htbp] 
  \centering
\subfigure[]{\label{fig:resonances_2a}
  \includegraphics[width=8cm]{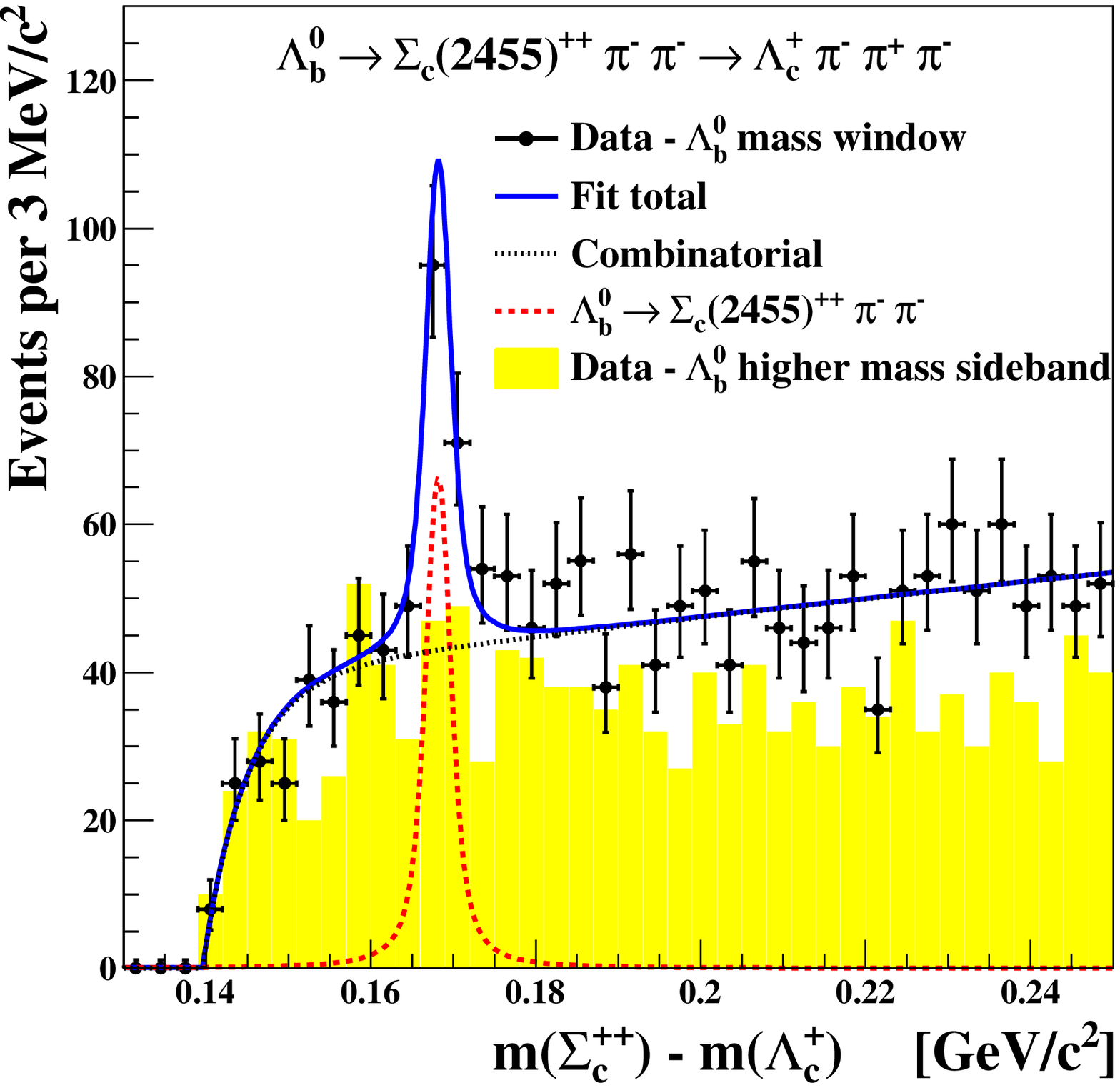}
}
\subfigure[]{\label{fig:resonances_2b}
  \includegraphics[width=8cm]{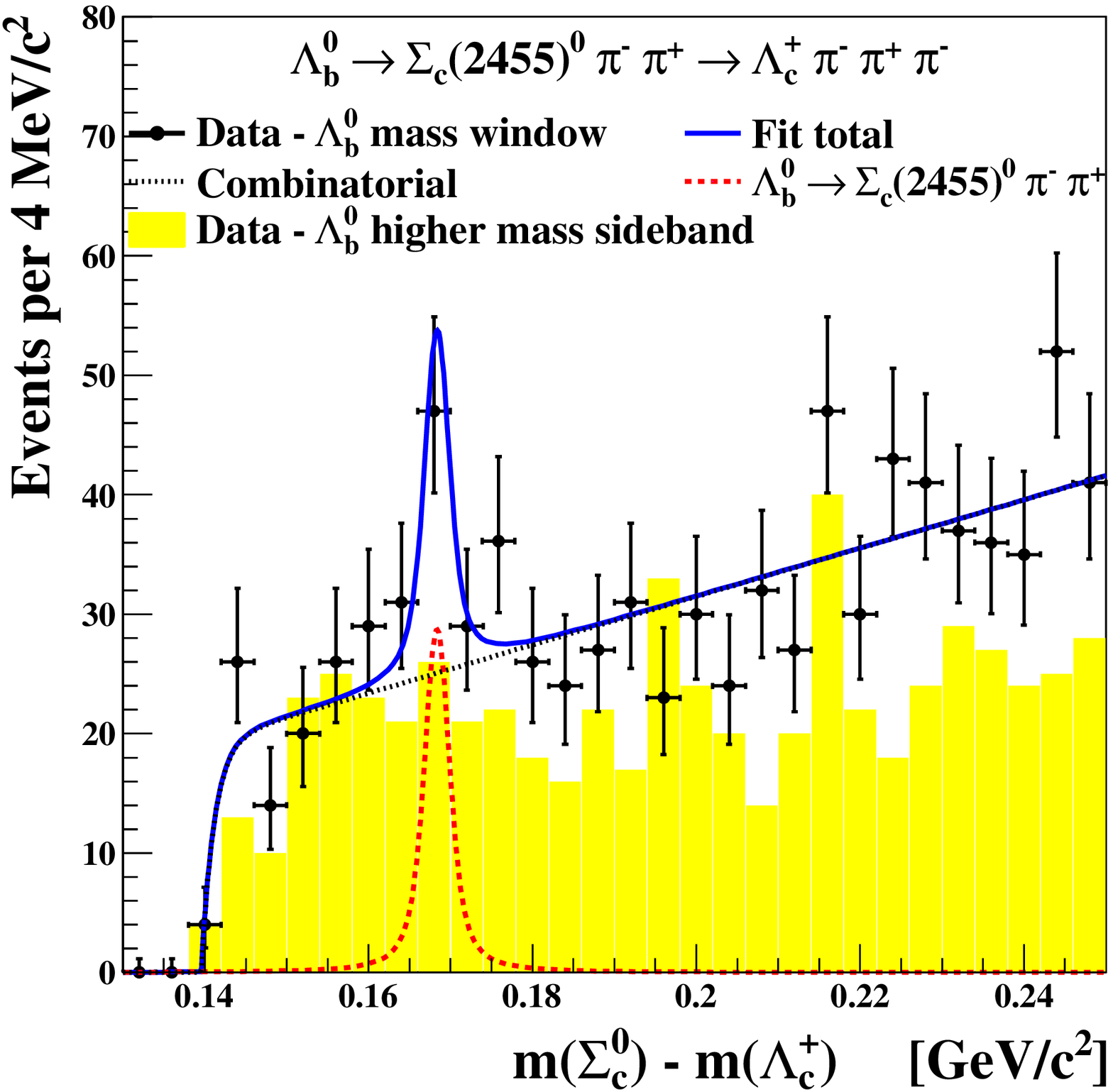}
}
\subfigure[]{\label{fig:resonances_2c}
  \includegraphics[width=8cm]{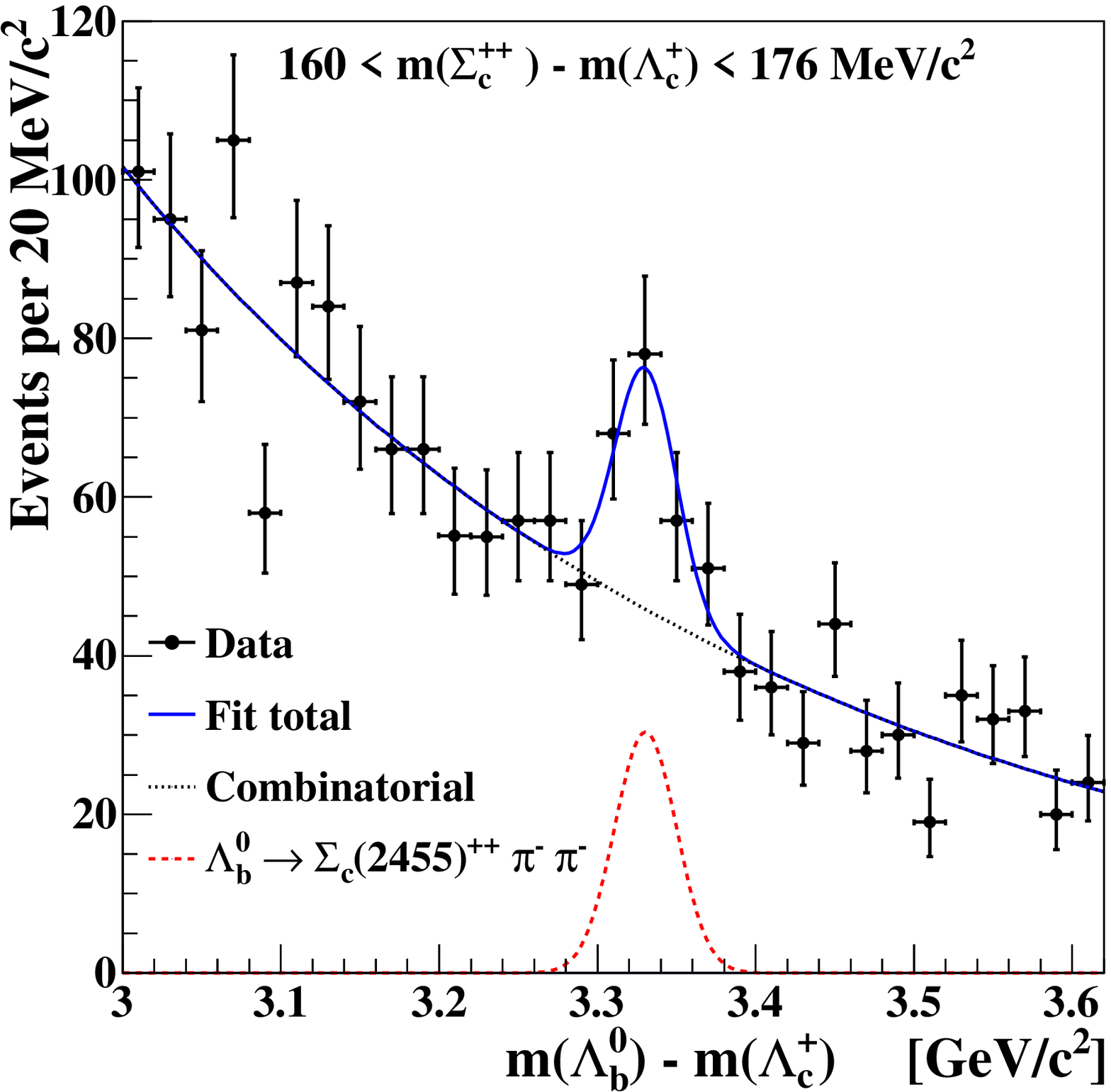}
}
\subfigure[]{\label{fig:resonances_2d}
  \includegraphics[width=8cm]{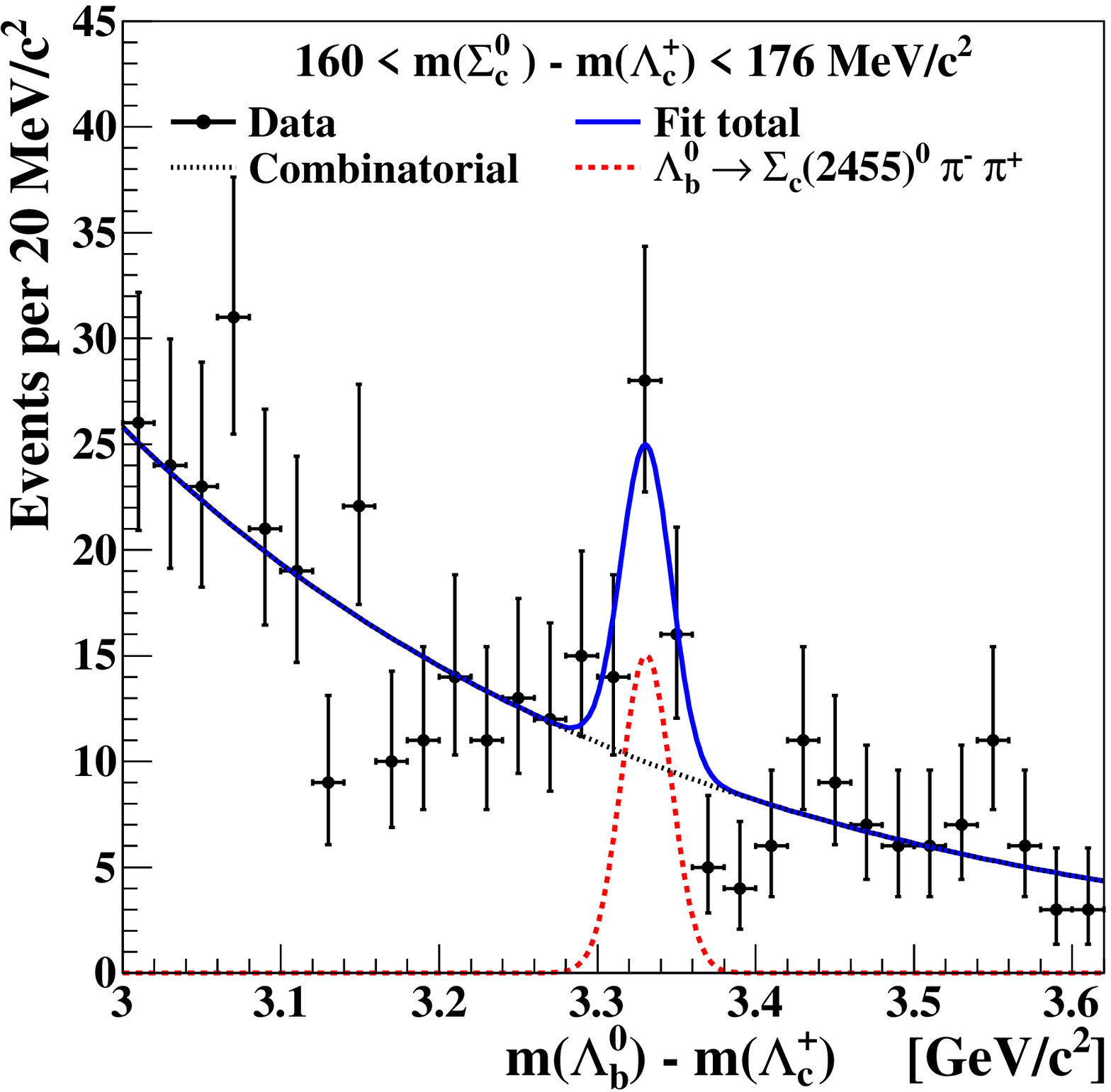}
}
  \caption{The $\Lambda^0_b\rightarrow\Sigma_c(2455)^{++}\pi^-\pi^-$
    and $\Lambda^0_b\rightarrow\Sigma_c(2455)^0\pi^+\pi^-$ signals:
    (a) $m(\Sigma_c(2455)^{++})-m(\Lambda^+_c)$ distribution 
for candidates in a
$\pm3\sigma$ range ($\pm$57 MeV/c$^2$) around the $\Lambda^0_b$ mass;
    (b) $m(\Sigma_c(2455)^0)-m(\Lambda^+_c)$ distribution 
for candidates in a
$\pm3\sigma$ range around the $\Lambda^0_b$ mass;
    (c) $m(\Lambda^0_b)-m(\Lambda^+_c)$ distribution 
restricted to candidates in the region 0.160
$< m(\Sigma_c(2455)^{++})-m(\Lambda^+_c)< 0.176$ GeV/c$^2$;
    (d)  $m(\Lambda^0_b)-m(\Lambda^+_c)$ distribution
restricted to candidates in the region 0.160 
$< m(\Sigma_c(2455)^0)-m(\Lambda^+_c)< 0.176$ GeV/c$^2$.
}
\label{fig:resonances_2}
\end{figure*}

To extract the $\Lambda^0_b\rightarrow\Sigma_c(2455)^{++}\pi^-\pi^-$
and $\Lambda^0_b\rightarrow\Sigma_c(2455)^{0}\pi^+\pi^-$ signals, 
the contributions due to the $\Lambda^0_b\rightarrow\Lambda_c(2595)^+\pi^-$
and $\Lambda^0_b\rightarrow\Lambda_c(2625)^+\pi^-$ decay modes
are removed by applying the veto requirement
$m(\Lambda_c^{*+})-m(\Lambda^+_c)>0.380$ GeV/c$^2$.
In Fig.~\ref{fig:resonances_2}\subref{fig:resonances_2a}
and \ref{fig:resonances_2}\subref{fig:resonances_2b} 
the resulting $m(\Sigma_c(2455)^{++})-m(\Lambda^+_c)$ and
$m(\Sigma_c(2455)^0)-m(\Lambda^+_c)$ distributions are shown.
Prominent $\Sigma_c(2455)^{++}$ and $\Sigma_c(2455)^0$ signals
are visible. 
While there is only one $\Sigma_c(2455)^{++}$ candidate 
for each
$\Lambda^0_b\rightarrow\Lambda^+_c\pi^-\pi^+\pi^-$ decay, two 
$\Sigma_c(2455)^0$ candidates are possible. 
Also in this case, only the candidate made with the
$\pi^-$ with lower $p_T$ is
in the $\Sigma_c(2455)^0$ mass region.
The $\Sigma_c(2455)^{++}$ and $\Sigma_c(2455)^0$ signals
are modeled with non-relativistic Breit-Wigner functions 
convolved with a Gaussian resolution function,
with the addition of an empirical background
\cite{Brun:1997pa, Antcheva:2009zz}. 
The $\Sigma_c(2455)^{++}$ and $\Sigma_c(2455)^0$  natural widths 
are Gaussian constrained to the world average values \cite{PDG2010}, 
while the width of the Gaussian resolution 
function is determined to be 1 MeV/c$^2$ from larger statistics samples of
$\Sigma_c(2455)^{++}$ and $\Sigma_c(2455)^0$ 
in the $\Lambda^0_b$ lower mass region and is fixed in the
fit. The effect of this approximation is taken into account
in the systematic uncertainties. The estimated 
$\Lambda^0_b\rightarrow\Sigma_c(2455)^{++}\pi^-\pi^-$ and 
$\Lambda^0_b\rightarrow\Sigma_c(2455)^0\pi^+\pi^-$ yields
and significances are reported in Table~\ref{tab:yields}.

In Fig.~\ref{fig:resonances_2}\subref{fig:resonances_2c}
and \ref{fig:resonances_2}\subref{fig:resonances_2d} 
the $m(\Lambda^0_b)-m(\Lambda^+_c)$ distributions are shown 
restricted to candidates with
0.160 $<m(\Sigma_c(2455)^{++,0})-m(\Lambda^+_c)< 0.176$ GeV/c$^2$, 
where the  $\Sigma_c(2455)^{++}$ and $\Sigma_c(2455)^0$ signals are
contained. The $\Lambda^0_b$ signal is modeled
with a Gaussian distribution, with floating mean and width, 
while the combinatorial background is  
an exponential function with floating slope and normalization.
We verified that the 
$\Lambda^0_b\rightarrow\Sigma_c(2455)^{++}\pi^-\pi^-$ and 
$\Lambda^0_b\rightarrow\Sigma_c(2455)^0\pi^+\pi^-$ yields
estimated by fitting the $m(\Lambda^0_b)-m(\Lambda^+_c)$ distributions
are compatible with the yields reported in Table~\ref{tab:yields}
with lower significance.
The fitted masses and widths of the four resonances are
in agreement with the world averages \cite{PDG2010}
and the recent CDF II measurements \cite{prd_cbaryons}.

\begin{table}[htb]
 \caption{Yields and significances of the 
$\Lambda^0_b\rightarrow\Lambda^+_c\pi^-\pi^+\pi^-$ decay modes. The quoted uncertainty 
is statistical only.}
  \begin{center}
    \begin{tabular}{l c c}
      \hline\hline
      \footnotesize{$\Lambda^0_b$ decay mode} & 
      \footnotesize{Yield}  &
\footnotesize{Significance($\sigma$)}\\
      \hline
      $\Lambda_c(2595)^{+}\pi^-\rightarrow\Lambda^+_c\pi^-\pi^+\pi^-$  & 
$46.0\pm8.2$ & 6.2 \\
      $\Lambda_c(2625)^{+}\pi^-\rightarrow\Lambda^+_c\pi^-\pi^+\pi^-$  & 
$135\pm15$ & $>$8 \\
      $\Sigma_c(2455)^{++}\pi^-\pi^-\rightarrow\Lambda^+_c\pi^-\pi^+\pi^-$  & 
$110\pm19$ & 6.6 \\
      $\Sigma_c(2455)^{0}\pi^+\pi^-\rightarrow\Lambda^+_c\pi^-\pi^+\pi^-$  & 
$36 \pm 11$ & 3.4 \\
      $\Lambda^+_c\pi^-\pi^+\pi^-(\rm{other})$  & $790 \pm 100$ & $>$8\\
      \hline\hline
    \end{tabular}
    \label{tab:yields}
  \end{center}
\end{table}

\begin{figure}[htbp] 
\begin{center}
\includegraphics[width=8cm]{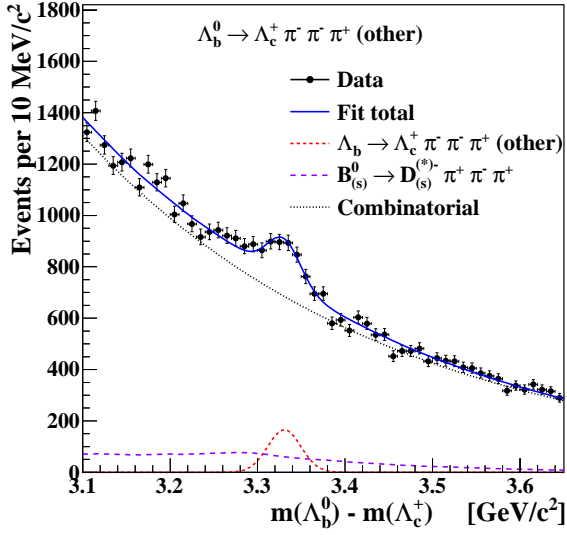}
\caption{The $\Lambda^0_b\rightarrow\Lambda^+_c\pi^-\pi^+\pi^-(\rm{other})$ signal
after vetoing
the resonant decay modes: $m(\Lambda^0_b)-m(\Lambda^+_c)$
distribution.}
\label{fig:LbLc3pi_other}
\end{center}
\end{figure}

The residual $\Lambda^0_b$ signal
(named $\Lambda^0_b\rightarrow\Lambda^+_c\pi^-\pi^+\pi^-(\rm{other})$)
is selected by applying the cuts 
$m(\Lambda_c^{*+})-m(\Lambda^+_c)>0.380$ GeV/c$^2$ and
$m(\Sigma_c(2455)^{++,0})-m(\Lambda^+_c)>0.190$ GeV/c$^2$ to
remove the contribution due to the resonant decay modes 
(Fig.~\ref{fig:LbLc3pi_other}).
This residual $\Lambda^0_b$ signal
is likely due to a combination of the
$\Lambda^0_b\rightarrow\Lambda^+_c a_1(1260)^-$, 
$\Lambda^0_b\rightarrow\Lambda^+_c\rho^0\pi^-$ with non-resonant
$\rho^0\pi^-$ (i.e. not produced by a $a_1(1260)^-$ decay), 
and non-resonant $\Lambda^0_b\rightarrow\Lambda^+_c\pi^-\pi^+\pi^-$
decay modes, in unknown proportions. 
A fit is performed with a Gaussian function, with floating
mean and width to model the signal, an exponential function
with floating slope and normalization to model 
the combinatorial background, and a physics background
due to the $B^0_{(s)}\rightarrow D_{(s)}^{(*)-}\pi^+\pi^-\pi^+$
decay modes, derived from simulation and included in the fit
with fixed shape and floating normalization. The resulting yield
is $790\pm$100 candidates (Table \ref{tab:yields}).
The unknown composition of the 
$\Lambda^0_b\rightarrow\Lambda^+_c\pi^-\pi^+\pi^-(\rm{other})$ sample
is taken into account as a source of systematic uncertainty.

\begin{table*}[htb]
 \caption{Measured branching fractions relative
to the $\Lambda^0_b\rightarrow\Lambda^+_c\pi^-$ decay mode
(second column). Absolute branching fractions (third column) 
are derived by normalizing
to the known value
${\mathcal{B}}(\Lambda^0_b\rightarrow\Lambda^+_c\pi^-)
=(8.8\pm3.2)\times10^{-3}$ \cite{prl_Lcpi}.
The first quoted uncertainty is statistical,
the second is systematic, and the third is due to the
uncertainty on the $\Lambda^0_b\rightarrow\Lambda^+_c\pi^-$
branching fraction.}
  \begin{center} 
    \begin{tabular}{l l l}
      \hline\hline 
\footnotesize{$\Lambda^0_b$ decay mode}
&
      \footnotesize{  
Relative {$\mathcal{B}$} to $\Lambda^0_b\rightarrow\Lambda^+_c\pi^-$} & 
\footnotesize{Absolute {$\mathcal{B}(10^{-3})$}}  \\
      \hline
 $\BR(\Lambda^0_b\rightarrow \Lambda_c(2595)^+\pi^-)\cdot
\BR(\Lambda_c(2595)^+\rightarrow\Lambda^+_c \pi^+\pi^-)$ 
&$  (7.1\pm1.3\pm0.6)\cdot10^{-2}$ & $0.62\pm0.11\pm0.05 \pm0.23$ \\

 $\BR(\Lambda^0_b\rightarrow \Lambda_c(2625)^+\pi^-)\cdot
\BR(\Lambda_c(2625)^+\rightarrow\Lambda^+_c \pi^+\pi^-)$
&$  (20.6\pm2.4^{+1.4}_{-1.5})\cdot10^{-2}$ & 
$1.81\pm0.21^{+0.12}_{-0.13}\pm0.66$ \\

$\BR(\Lambda^0_b\rightarrow \Sigma_c(2455)^{++}\pi^- \pi^-)\cdot
\BR(\Sigma_c(2455)^{++}\rightarrow\Lambda^+_c \pi^+)$
&$  (19.0\pm3.3\pm1.1)\cdot10^{-2}$ & 
$1.67\pm0.29\pm0.10\pm0.61$ \\

$\BR(\Lambda^0_b\rightarrow \Sigma_c(2455)^{0}\pi^+ \pi^-)\cdot
\BR(\Sigma_c(2455)^0\rightarrow\Lambda^+_c \pi^-)$
&$  (21.5\pm6.5^{+4.5}_{-2.9})\cdot10^{-2}$ & 
$1.89\pm0.57^{+0.40}_{-0.26}\pm0.69$ \\

$\BR(\Lambda^0_b\rightarrow\Lambda^+_c\pi^-\pi^+\pi^-(\rm{other}))$
&$  2.36\pm0.32^{+0.68}_{-0.53}$ &
$20.8\pm2.8^{+6.0}_{-4.7}\pm7.6$ \\

$\BR(\Lambda^0_b\rightarrow\Lambda^+_c\pi^-\pi^+\pi^-)$
&$  3.04\pm 0.33^{+0.70}_{-0.55}$ & 
$26.8\pm2.9^{+6.2}_{-4.8}\pm 9.7$ \\
      \hline\hline
    \end{tabular}
    \label{tab:measure}
  \end{center}
\end{table*}
\section{Measurement of the ratio of branching fractions
${\mathcal{B}}(\Lambda^0_b\rightarrow \Lambda^+_c\pi^-\pi^+\pi^-)
/{\mathcal{B}}(\Lambda^0_b\rightarrow \Lambda^+_c\pi^-)$}
\label{sec:measurement}
We measure the following ratio of branching fractions:
\begin{displaymath}
\frac{\BR(\Lambda^0_b\rightarrow\Lambda^+_c\pi^-\pi^+\pi^-)}
{\BR(\Lambda^0_b\rightarrow\Lambda^+_c \pi^-)} =
\end{displaymath}
\begin{equation}
=\sum_i\frac{N(\Lambda^0_b\rightarrow i \rightarrow\Lambda^+_c\pi^-\pi^+\pi^-)
}
{N(\Lambda^0_b\rightarrow\Lambda^+_c \pi^-)
}
\frac{\epsilon_{\Lambda^0_b\rightarrow\Lambda^+_c\pi^-}}
{\epsilon_i},
\nonumber
\end{equation}
where $N$ are the measured signal yields reported in Table \ref{tab:yields}, 
and the sum on the intermediate ``$i$" states includes
$\Lambda_c(2595)^+\pi^-$, $\Lambda_c(2625)^+\pi^-$,  $\Sigma_c(2455)^{++}\pi^-\pi^-$,  
$\Sigma_c(2455)^{0}\pi^+\pi^-$, and $\Lambda^+_c\pi^-\pi^+\pi^-(\rm{other})$.
In the last state, 
we assume equal proportions of 
the three decay modes 
$\Lambda^0_b\rightarrow\Lambda^+_c a_1(1260)^-$,
$\Lambda^0_b\rightarrow\Lambda^+_c\rho^0\pi^-$, and
non-resonant $\Lambda^0_b\rightarrow\Lambda^+_c\pi^-\pi^+\pi^-$.
To convert event yields into relative branching fractions,
we apply the corrections 
$\epsilon_{\Lambda^0_b\rightarrow\Lambda^+_c\pi^-}/\epsilon_i$
for the various
trigger and offline selection efficiencies of
the decay modes $\Lambda^0_b\rightarrow\Lambda^+_c \pi^-$
and $\Lambda^0_b\rightarrow i\rightarrow\Lambda^+_c\pi^-\pi^+\pi^-$. 
All corrections are determined
from the detailed detector simulation.
The {\sc bgenerator} program produces samples of specific
$B$ hadron decays according to measured $p_T$ and rapidity
spectra \cite{BGen}. 
Decays of $b$ and $c$ hadrons and their
daughters are simulated using the {\sc evtgen} 
package \cite{evtgen}.
The geometry and response of the detector components 
are simulated with the {\sc geant} software package \cite{geant}
and simulated events are processed with a full
simulation of the CDF~II detector and trigger.
The resulting estimated corrections
$\epsilon_{\Lambda^0_b\rightarrow\Lambda^+_c\pi^-}/\epsilon_i$
are $4.70 \pm 0.10$, $4.66 \pm 0.10$, $5.28 \pm 0.11$, and $18.49 \pm 0.66$, respectively, for the   
$\Lambda_c(2595)^+\pi^-$, 
$\Lambda_c(2625)^+\pi^-$, 
$\Sigma_c(2455)^{++}\pi^-\pi^-$, and 
$\Sigma_c(2455)^{0}\pi^+\pi^-$ decay modes.
For the $\Lambda^+_c\pi^-\pi^+\pi^-(\rm{other})$ decay mode a
correction factor equal to $9.16 \pm 0.14$ is obtained by averaging the 
relative efficiencies of the three intermediate states
$\Lambda^0_b\rightarrow\Lambda^+_c a_1(1260)^-$,
$\Lambda^0_b\rightarrow\Lambda^+_c\rho^0\pi^-$, and
non-resonant $\Lambda^0_b\rightarrow\Lambda^+_c\pi^-\pi^+\pi^-$.\\
With a similar method, we also measure the ratios of the branching fractions 
of the intermediate resonances contributing 
to $\Lambda^0_b\rightarrow\Lambda^+_c\pi^-\pi^+\pi^-$,
\begin{displaymath}
\frac{\BR(\Lambda^0_b\rightarrow j \rightarrow\Lambda^+_c\pi^-\pi^+\pi^-)}
{\BR(\Lambda^0_b\rightarrow\Lambda^+_c\pi^-\pi^+\pi^-)}=
\end{displaymath}
\begin{equation}
=\frac{N(\Lambda^0_b\rightarrow j \rightarrow\Lambda^+_c\pi^-\pi^+\pi^-)}{\sum_{i}N(\Lambda^0_b\rightarrow i \rightarrow\Lambda^+_c\pi^-\pi^+\pi^-) \frac{\epsilon_j}{\epsilon_i} }.
\nonumber
\end{equation}
\section{Systematic uncertainties}
\label{sec:systematics}
\begin{figure}[htbp] 
\begin{center}
\includegraphics[width=8cm]{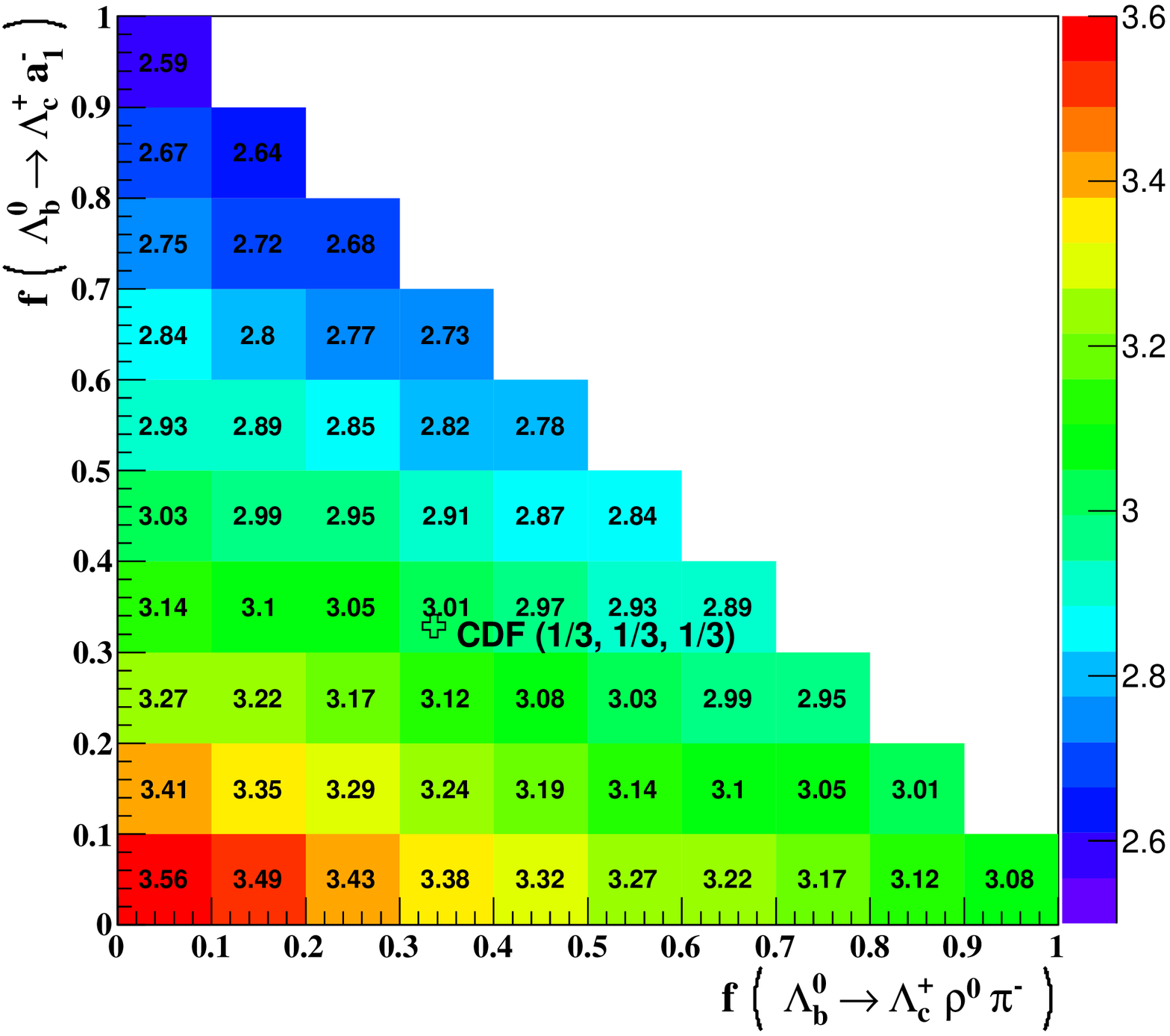}
\caption{
$\BR(\Lambda^0_b\rightarrow \Lambda^+_c\pi^-\pi^+\pi^-)
/\BR(\Lambda^0_b\rightarrow \Lambda^+_c \pi^-)$ (color scale) 
as a function of the assumed fractions of
$\Lambda^0_b\rightarrow \Lambda^+_c a^-_1$ and
$\Lambda^0_b\rightarrow \Lambda^+_c \rho^0 \pi^-$
 in the composition of the $\Lambda^0_b\rightarrow\Lambda^+_c\pi^-\pi^+\pi^-(\rm{other})$ sample.
The central value of the ratio is overlaid in each bin.
The fraction of non-resonant $\Lambda^0_b\rightarrow \Lambda^+_c\pi^-\pi^+\pi^-$
is equal to 
$1 - f\left(\Lambda^0_b\rightarrow \Lambda^+_c a^-_1\right) 
- f\left(\Lambda^0_b\rightarrow \Lambda^+_c \rho^0 \pi^-\right)$. 
The cross represents the composition chosen for the present
measurement assuming equal proportions of 
$\Lambda^0_b\rightarrow \Lambda^+_c a^-_1$,   
$\Lambda^0_b\rightarrow \Lambda^+_c \rho^0 \pi^-$ 
and non-resonant $\Lambda^0_b\rightarrow \Lambda^+_c\pi^-\pi^+\pi^-$.
}
\label{color}
\end{center}
\end{figure}

The dominant sources of systematic uncertainty are
the unknown relative fractions of $\Lambda^0_b\rightarrow\Lambda^+_c a_1(1260)^-$,
$\Lambda^0_b\rightarrow\Lambda^+_c \rho^0\pi^-$, and
non-resonant $\Lambda^0_b\rightarrow\Lambda^+_c\pi^-\pi^+\pi^-$,
which affect 
the $\Lambda^0_b\rightarrow\Lambda^+_c\pi^-\pi^+\pi^-(\rm{other})$ decay mode efficiency, and
the unknown $\Lambda^0_b$ production 
and $\Lambda^+_c$ decay polarizations, which affect
the estimate of all the $\epsilon_i$
and $\epsilon_{\Lambda^0_b\rightarrow\Lambda^+_c\pi^-}$ efficiencies.
The correction
$\epsilon_{\Lambda^0_b\rightarrow\Lambda^+_c\pi^-}/\epsilon_{\Lambda^0_b\rightarrow\Lambda^+_c\pi^-\pi^+\pi^-(\rm{other})}$ 
has an average value of 9.16 and varies
between a minimum of 7.4 and a maximum of 11.6, 
obtained in the extreme cases in which the 
$\Lambda^0_b\rightarrow\Lambda^+_c\pi^-\pi^+\pi^-(\rm{other})$
sample is assumed to be entirely composed of 
$\Lambda^0_b\rightarrow \Lambda^+_c a_1(1260)^-$ or non-resonant 
$\Lambda^0_b\rightarrow \Lambda^+_c\pi^-\pi^+\pi^-$, respectively.
The dependence of
$\BR(\Lambda^0_b\rightarrow \Lambda^+_c\pi^-\pi^+\pi^-)
/\BR(\Lambda^0_b\rightarrow \Lambda^+_c \pi^-)$  
on the fraction of $\Lambda^0_b\rightarrow \Lambda^+_c a_1(1260)^-$ and
$\Lambda^0_b\rightarrow \Lambda^+_c \rho^0 \pi^-$ in 
 the $\Lambda^0_b\rightarrow\Lambda^+_c\pi^-\pi^+\pi^-(\rm{other})$ sample 
is shown in Fig.~\ref{color}.
The difference between the values computed with the average and 
the minimum (maximum) efficiency correction, respectively,
is taken as an estimate of the lower (upper) associated 
systematic uncertainty.\\
The unpolarized $\Lambda^0_b$ and $\Lambda^+_c$
simulation samples are used to obtain the central
values of the efficiency corrections. For the study of
the systematic uncertainties, angular distributions in 
simulation are reweighted according to all possible 
combinations of the $\Lambda^0_b$ production polarization states
along the normal to the production plane,
with the $\Lambda^+_c$ polarization states.
The $\Lambda^0_b$ polarization and the $\Lambda^+_c$
polarizations are both taken to vary independently in the range 
$\pm1$. We assume the extreme scenarios where both the
$\Lambda^0_b$ and $\Lambda^+_c$ baryons are 100\% polarized
and we recompute the efficiency corrections assuming the four
possible $\Lambda^0_b$ and $\Lambda^+_c$ polarization
combinations.
The difference in the efficiency corrections between
the simulation with reweighted angular distributions
and the simulation with unpolarized $\Lambda^0_b$ and
$\Lambda^+_c$ is used to determine the associated
systematic uncertainty. These two sources of systematic
uncertainty account for approximately 98\% of the total 
systematic uncertainty on the measurement of the relative
branching fraction
$\BR(\Lambda^0_b\rightarrow \Lambda^+_c\pi^-\pi^+\pi^-)
/\BR(\Lambda^0_b\rightarrow \Lambda^+_c \pi^-)$. 
Other systematic errors stem from the uncertainties on the 
$\Lambda^0_b\rightarrow\Lambda^+_c\pi^-$ background shapes;
on the Cabibbo suppressed decay mode contributions,
which affect the estimate of the signal yields;
on the Monte Carlo simulation of the signal decay modes
(limited sample statistics, trigger emulation, and 
$\Lambda^0_b$ production transverse momentum distribution),
which affect the estimate of the efficiency corrections.
The contributions due to the uncertainties on 
the $\Sigma_c^{++}$ and $\Sigma^0_c$ signal and
background shapes, the $\Lambda^+_c$ and  $\Lambda_c^{*+}$ branching fractions,
and  the $\Lambda^0_b$ and $\Lambda^+_c$ 
lifetimes are negligible.

As a cross-check of the analysis, 
we also measure the relative branching fraction  
$\BR(B^0\rightarrow D^-\pi^+\pi^-\pi^+)/\BR(B^0\rightarrow D^-\pi^+)$,
using the same data sample and  vertex
reconstruction procedure developed for the $\Lambda^0_b$ analysis.
We apply the same optimized cuts to the $B^0$ candidates,  
with the additional request to have a $D^-$ candidate with mass
within $\pm$22 MeV/c$^2$ of the known mass of $D^-$ \cite{PDG2010}.
We estimate  $B^0\rightarrow D^-\pi^+\pi^-\pi^+$ and  $B^0\rightarrow D^-\pi^+$ yields
of $431 \pm 32$ and $1352 \pm 44$ candidates, respectively.
Our measurement 
$\BR(B^0\rightarrow D^-\pi^+\pi^-\pi^+)/\BR(B^0\rightarrow D^-\pi^+)
= 3.06 \pm 0.25(\text{stat})$
is in good agreement with the value calculated from
the measured absolute branching fractions of the 
$B^0$ decay modes reported in Ref.~\cite{PDG2010}.
\section{Results}
\label{sec:results}
We measure the relative branching ratio of 
$\Lambda^0_b\rightarrow \Lambda^+_c\pi^-\pi^+\pi^-$ to $\Lambda^0_b\rightarrow \Lambda^+_c\pi^-$ decays to be
\begin{displaymath}
\frac{\BR(\Lambda^0_b\rightarrow \Lambda^+_c \pi^-\pi^+\pi^-)}{\BR(\Lambda^0_b\rightarrow \Lambda^+_c \pi^-)} = 3.04 \pm 0.33(\text{stat})^{+0.70}_{-0.55}(\text{syst}).
\end{displaymath}
The relative branching fractions of the intermediate states contributing to $\Lambda^0_b\rightarrow \Lambda^+_c \pi^-\pi^+\pi^-$
with respect to $\Lambda^0_b\rightarrow \Lambda^+_c\pi^-$ are reported in Table \ref{tab:measure}. 
The absolute branching fractions are derived by normalizing
to the known value ${\mathcal{B}}(\Lambda^0_b\rightarrow\Lambda^+_c\pi^-) = (8.8\pm3.2)\times10^{-3}$ \cite{prl_Lcpi}.\\
To compare our result with the recent LHCb measurement \cite{LHCb} of
$1.43 \pm 0.16(\text{stat})\pm 0.13(\text{syst})$,  
we assume the composition of the admixture
to be 2/3 $\Lambda^0_b\rightarrow\Lambda^+_c a_1(1260)^-$ and
1/3 $\Lambda^0_b\rightarrow\Lambda^+_c \rho^0\pi^-$, 
and use the overall $\Lambda^0_b\rightarrow \Lambda^+_c\pi^-\pi^+\pi^-$
yield and a global efficiency correction
to compute $\BR(\Lambda^0_b\rightarrow \Lambda^+_c\pi^-\pi^+\pi^-)/\BR(\Lambda^0_b\rightarrow \Lambda^+_c \pi^-)$, as 
in the LHCb analysis.
This results in a value of $2.55 \pm 0.25(\text{stat})^{+0.82}_{-0.27}(\text{syst})$, 
which is inconsistent with the 
LHCb result at the level of 2.6 Gaussian standard deviations.\\
We also measure the relative branching fractions of the intermediate resonances 
contributing to the $\Lambda^0_b\rightarrow\Lambda^+_c\pi^-\pi^+\pi^-$ decay (Table~\ref{tab:measure_1}). 
These results are of comparable or higher precision than
existing measurements.
\begin{table*}[htb]
 \caption{Measured branching fractions of the resonant decay modes relative
to $\Lambda^0_b\rightarrow\Lambda^+_c\pi^-\pi^+\pi^-$. 
The first quoted uncertainty is statistical,
the second is systematic.}
  \begin{center} 
    \begin{tabular}{l l}
      \hline\hline 
\footnotesize{$\Lambda^0_b$ decay mode}
&
      \footnotesize{  
Relative {$\mathcal{B}(10^{-2})$}} \\
      \hline
 $\BR(\Lambda^0_b\rightarrow \Lambda_c(2595)^+\pi^-)\cdot
\BR(\Lambda_c(2595)^+\rightarrow\Lambda^+_c \pi^+\pi^-)$ 
&$  2.3\pm0.5 \pm0.4$ \\

 $\BR(\Lambda^0_b\rightarrow \Lambda_c(2625)^+\pi^-)\cdot
\BR(\Lambda_c(2625)^+\rightarrow\Lambda^+_c \pi^+\pi^-)$
&$  6.8\pm1.0\pm1.3$ \\

$\BR(\Lambda^0_b\rightarrow \Sigma_c(2455)^{++}\pi^- \pi^-)\cdot
\BR(\Sigma_c(2455)^{++}\rightarrow\Lambda^+_c \pi^+)$
&$  6.2\pm1.2\pm1.3$ \\

$\BR(\Lambda^0_b\rightarrow \Sigma_c(2455)^{0}\pi^+ \pi^-)\cdot
\BR(\Sigma_c(2455)^0\rightarrow\Lambda^+_c \pi^-)$
&$  7.1\pm2.1^{+1.5}_{-1.3}$ \\

$\BR(\Lambda^0_b\rightarrow\Lambda^+_c\pi^-\pi^+\pi^-(\rm{other}))$
&$  77.6\pm3.0^{+4.0}_{-4.1}$\\

      \hline\hline
    \end{tabular}
    \label{tab:measure_1}
  \end{center}
\end{table*}

\section{Conclusion}
\label{sec:conclusion}
In summary, we reconstruct the 
$\Lambda^0_b\rightarrow\Lambda^+_c\pi^-\pi^+\pi^-$
decay mode and the 
$\Lambda^0_b\rightarrow\Lambda_c(2595)^+\pi^-$,
$\Lambda^0_b\rightarrow\Lambda_c(2625)^+\pi^-$,
$\Lambda^0_b\rightarrow\Sigma_c(2455)^{++}\pi^-\pi^-$,
and $\Lambda^0_b\rightarrow\Sigma_c(2455)^{0}\pi^+\pi^-$
resonant decay modes in CDF~II data corresponding to 
2.4 fb$^{-1}$ of integrated luminosity.
We measure the branching fraction of the resonant 
decay modes relative to the $\Lambda^0_b\rightarrow\Lambda^+_c\pi^-$
branching fraction. We also measure
$\BR(\Lambda^0_b\rightarrow \Lambda^+_c\pi^-\pi^+\pi^-)
/\BR(\Lambda^0_b\rightarrow \Lambda^+_c \pi^-) 
= 3.04 \pm 0.33(\text{stat})^{+0.70}_{-0.55}(\text{syst})$.
Using the known value of 
$\BR(\Lambda^0_b\rightarrow\Lambda^+_c\pi^-)$ \cite{prl_Lcpi}
we  find
$\BR(\Lambda^0_b\rightarrow \Lambda^+_c\pi^-\pi^+\pi^-)=
(26.8 \pm 2.9(\text{stat})^{+6.2}_{-4.8}(\text{syst})\pm 9.7 (\text{norm}))\times 10^{-3}$,
where the third quoted uncertainty arises from the $\Lambda^0_b\rightarrow\Lambda^+_c\pi^-$ normalization uncertainty.\\
\section{Acknowledgments}
We thank the Fermilab staff and the technical staffs of the participating institutions for their vital contributions. This work was supported by the U.S. Department of Energy and National Science Foundation; the Italian Istituto Nazionale di Fisica Nucleare; the Ministry of Education, Culture, Sports, Science and Technology of Japan; the Natural Sciences and Engineering Research Council of Canada; the National Science Council of the Republic of China; the Swiss National Science Foundation; the A.P. Sloan Foundation; the Bundesministerium f\"ur Bildung und Forschung, Germany; the Korean World Class University Program, the National Research Foundation of Korea; the Science and Technology Facilities Council and the Royal Society, UK; the Russian Foundation for Basic Research; the Ministerio de Ciencia e Innovaci\'{o}n, and Programa Consolider-Ingenio 2010, Spain; the Slovak R\&D Agency; the Academy of Finland; and the Australian Research Council (ARC).

 \end{document}

%% file: September2011_Authors.tex
\affiliation{Institute of Physics, Academia Sinica, Taipei, Taiwan 11529, Republic of China}
\affiliation{Argonne National Laboratory, Argonne, Illinois 60439, USA}
\affiliation{University of Athens, 157 71 Athens, Greece}
\affiliation{Institut de Fisica d'Altes Energies, ICREA, Universitat Autonoma de Barcelona, E-08193, Bellaterra (Barcelona), Spain}
\affiliation{Baylor University, Waco, Texas 76798, USA}
\affiliation{Istituto Nazionale di Fisica Nucleare Bologna, $^{cc}$University of Bologna, I-40127 Bologna, Italy}
\affiliation{University of California, Davis, Davis, California 95616, USA}
\affiliation{University of California, Los Angeles, Los Angeles, California 90024, USA}
\affiliation{Instituto de Fisica de Cantabria, CSIC-University of Cantabria, 39005 Santander, Spain}
\affiliation{Carnegie Mellon University, Pittsburgh, Pennsylvania 15213, USA}
\affiliation{Enrico Fermi Institute, University of Chicago, Chicago, Illinois 60637, USA}
\affiliation{Comenius University, 842 48 Bratislava, Slovakia; Institute of Experimental Physics, 040 01 Kosice, Slovakia}
\affiliation{Joint Institute for Nuclear Research, RU-141980 Dubna, Russia}
\affiliation{Duke University, Durham, North Carolina 27708, USA}
\affiliation{Fermi National Accelerator Laboratory, Batavia, Illinois 60510, USA}
\affiliation{University of Florida, Gainesville, Florida 32611, USA}
\affiliation{Laboratori Nazionali di Frascati, Istituto Nazionale di Fisica Nucleare, I-00044 Frascati, Italy}
\affiliation{University of Geneva, CH-1211 Geneva 4, Switzerland}
\affiliation{Glasgow University, Glasgow G12 8QQ, United Kingdom}
\affiliation{Harvard University, Cambridge, Massachusetts 02138, USA}
\affiliation{Division of High Energy Physics, Department of Physics, University of Helsinki and Helsinki Institute of Physics, FIN-00014, Helsinki, Finland}
\affiliation{University of Illinois, Urbana, Illinois 61801, USA}
\affiliation{The Johns Hopkins University, Baltimore, Maryland 21218, USA}
\affiliation{Institut f\"{u}r Experimentelle Kernphysik, Karlsruhe Institute of Technology, D-76131 Karlsruhe, Germany}
\affiliation{Center for High Energy Physics: Kyungpook National University, Daegu 702-701, Korea; Seoul National University, Seoul 151-742, Korea; Sungkyunkwan University, Suwon 440-746, Korea; Korea Institute of Science and Technology Information, Daejeon 305-806, Korea; Chonnam National University, Gwangju 500-757, Korea; Chonbuk National University, Jeonju 561-756, Korea}
\affiliation{Ernest Orlando Lawrence Berkeley National Laboratory, Berkeley, California 94720, USA}
\affiliation{University of Liverpool, Liverpool L69 7ZE, United Kingdom}
\affiliation{University College London, London WC1E 6BT, United Kingdom}
\affiliation{Centro de Investigaciones Energeticas Medioambientales y Tecnologicas, E-28040 Madrid, Spain}
\affiliation{Massachusetts Institute of Technology, Cambridge, Massachusetts 02139, USA}
\affiliation{Institute of Particle Physics: McGill University, Montr\'{e}al, Qu\'{e}bec, Canada H3A~2T8; Simon Fraser University, Burnaby, British Columbia, Canada V5A~1S6; University of Toronto, Toronto, Ontario, Canada M5S~1A7; and TRIUMF, Vancouver, British Columbia, Canada V6T~2A3}
\affiliation{University of Michigan, Ann Arbor, Michigan 48109, USA}
\affiliation{Michigan State University, East Lansing, Michigan 48824, USA}
\affiliation{Institution for Theoretical and Experimental Physics, ITEP, Moscow 117259, Russia}
\affiliation{University of New Mexico, Albuquerque, New Mexico 87131, USA}
\affiliation{The Ohio State University, Columbus, Ohio 43210, USA}
\affiliation{Okayama University, Okayama 700-8530, Japan}
\affiliation{Osaka City University, Osaka 588, Japan}
\affiliation{University of Oxford, Oxford OX1 3RH, United Kingdom}
\affiliation{Istituto Nazionale di Fisica Nucleare, Sezione di Padova-Trento, $^{dd}$University of Padova, I-35131 Padova, Italy}
\affiliation{University of Pennsylvania, Philadelphia, Pennsylvania 19104, USA}
\affiliation{Istituto Nazionale di Fisica Nucleare Pisa, $^{ee}$University of Pisa, $^{ff}$University of Siena and $^{gg}$Scuola Normale Superiore, I-56127 Pisa, Italy}
\affiliation{University of Pittsburgh, Pittsburgh, Pennsylvania 15260, USA}
\affiliation{Purdue University, West Lafayette, Indiana 47907, USA}
\affiliation{University of Rochester, Rochester, New York 14627, USA}
\affiliation{The Rockefeller University, New York, New York 10065, USA}
\affiliation{Istituto Nazionale di Fisica Nucleare, Sezione di Roma 1, $^{hh}$Sapienza Universit\`{a} di Roma, I-00185 Roma, Italy}
\affiliation{Rutgers University, Piscataway, New Jersey 08855, USA}
\affiliation{Texas A\&M University, College Station, Texas 77843, USA}
\affiliation{Istituto Nazionale di Fisica Nucleare Trieste/Udine, I-34100 Trieste, $^{ii}$University of Udine, I-33100 Udine, Italy}
\affiliation{University of Tsukuba, Tsukuba, Ibaraki 305, Japan}
\affiliation{Tufts University, Medford, Massachusetts 02155, USA}
\affiliation{University of Virginia, Charlottesville, Virginia 22906, USA}
\affiliation{Waseda University, Tokyo 169, Japan}
\affiliation{Wayne State University, Detroit, Michigan 48201, USA}
\affiliation{University of Wisconsin, Madison, Wisconsin 53706, USA}
\affiliation{Yale University, New Haven, Connecticut 06520, USA}

\author{T.~Aaltonen}
\affiliation{Division of High Energy Physics, Department of Physics, University of Helsinki and Helsinki Institute of Physics, FIN-00014, Helsinki, Finland}
\author{B.~\'{A}lvarez~Gonz\'{a}lez$^x$}
\affiliation{Instituto de Fisica de Cantabria, CSIC-University of Cantabria, 39005 Santander, Spain}
\author{S.~Amerio}
\affiliation{Istituto Nazionale di Fisica Nucleare, Sezione di Padova-Trento, $^{dd}$University of Padova, I-35131 Padova, Italy}
\author{D.~Amidei}
\affiliation{University of Michigan, Ann Arbor, Michigan 48109, USA}
\author{A.~Anastassov$^v$}
\affiliation{Fermi National Accelerator Laboratory, Batavia, Illinois 60510, USA}
\author{A.~Annovi}
\affiliation{Laboratori Nazionali di Frascati, Istituto Nazionale di Fisica Nucleare, I-00044 Frascati, Italy}
\author{J.~Antos}
\affiliation{Comenius University, 842 48 Bratislava, Slovakia; Institute of Experimental Physics, 040 01 Kosice, Slovakia}
\author{G.~Apollinari}
\affiliation{Fermi National Accelerator Laboratory, Batavia, Illinois 60510, USA}
\author{J.A.~Appel}
\affiliation{Fermi National Accelerator Laboratory, Batavia, Illinois 60510, USA}
\author{T.~Arisawa}
\affiliation{Waseda University, Tokyo 169, Japan}
% ...

\collaboration{CDF Collaboration\footnote{With visitors from
$^a$Istituto Nazionale di Fisica Nucleare, Sezione di Cagliari, 09042 Monserrato (Cagliari), Italy,
$^b$University of CA Irvine, Irvine, CA 92697, USA,
$^c$University of CA Santa Barbara, Santa Barbara, CA 93106, USA,
$^d$University of CA Santa Cruz, Santa Cruz, CA 95064, USA,
$^e$CERN, CH-1211 Geneva, Switzerland,
$^f$Cornell University, Ithaca, NY 14853, USA,
$^g$University of Cyprus, Nicosia CY-1678, Cyprus,
$^h$Office of Science, U.S. Department of Energy, Washington, DC 20585, USA,
$^i$University College Dublin, Dublin 4, Ireland,
$^j$ETH, 8092 Zurich, Switzerland,
$^k$University of Fukui, Fukui City, Fukui Prefecture, Japan 910-0017,
$^l$Universidad Iberoamericana, Mexico D.F., Mexico,
$^m$University of Iowa, Iowa City, IA 52242, USA,
$^n$Kinki University, Higashi-Osaka City, Japan 577-8502,
$^o$Kansas State University, Manhattan, KS 66506, USA,
$^p$University of Manchester, Manchester M13 9PL, United Kingdom,
$^q$Queen Mary, University of London, London, E1 4NS, United Kingdom,
$^r$University of Melbourne, Victoria 3010, Australia,
$^s$Muons, Inc., Batavia, IL 60510, USA,
$^t$Nagasaki Institute of Applied Science, Nagasaki, Japan,
$^u$National Research Nuclear University, Moscow, Russia,
$^v$Northwestern University, Evanston, IL 60208, USA,
$^w$University of Notre Dame, Notre Dame, IN 46556, USA,
$^x$Universidad de Oviedo, E-33007 Oviedo, Spain,
$^y$CNRS-IN2P3, Paris, F-75252 France,
$^z$Texas Tech University, Lubbock, TX 79609, USA,
$^{aa}$Universidad Tecnica Federico Santa Maria, 110v Valparaiso, Chile,
$^{bb}$Yarmouk University, Irbid 211-63, Jordan,
}}
\noaffiliation

%% file: text.bbl
\begin{thebibliography}{1000}
\bibitem{Pseudorapidity}The pseudorapidity  is defined as $\eta = -\log{\tan{\left(\theta/2\right)}}$
where $\theta$ is the angle between the trajectory of the particle being considered and the undeflected beam direction.
\bibitem{CDF-coordinates}CDF~II uses a cylindrical coordinate
system in which $\phi$ is the azimuthal angle, $r$ is the radius
from the nominal beam line, and $z$ points in the proton beam
direction, with the origin at the center of the detector. 
The transverse plane is the plane perpendicular to the $z$ axis.
\bibitem{chrg_cnj} Throughout this article, the inclusion of
charge conjugate decays is implied.
\end{thebibliography}
